\documentclass[9pt,twocolumn,twoside]{ArXiVjnl}

\journal{ol} 

\setboolean{shortarticle}{true}
\usepackage{multicol,blindtext,graphics,graphicx,float,multirow,array,makecell,fourier,subcaption,caption,makecell}

\usepackage{color}

\title{Improving the beam quality factor (M\textsuperscript{2}) by phase-only reshaping of structured light}

\author[1]{Stirling Scholes}
\author[1,*]{Andrew Forbes}
\affil[1]{University of the Witwatersrand, Structured Light Group, Physics, 1 Jan Smuts Avenue, Johannesburg, South Africa, 2000}

\affil[*]{Corresponding author: Andrew.Forbes@wits.ac.za}

\begin{abstract}
Laser brightness is crucial in many optical processes, and is optimised by high power, high beam quality (low $M^2$) beams.  Here we show how to improve the laser beam quality factor (reducing the $M^2$) of arbitrary structured light fields in a lossless manner using continuous phase-only elements, thus allowing for the increase in brightness by a simple linear optical transformation. We demonstrate the principle with four high $M^2$ initial beams, converting each to a Gaussian ($M^2 \approx 1$) with a dramatic increase in brightness of $>10 \times$.  This work puts a new perspective on the old debate of improving laser beam quality with binary diffractive optics, while providing a practical approach to enhancing laser brightness for arbitrary input beams.
\end{abstract}

\setboolean{displaycopyright}{true}

\begin{document}

\maketitle
\begin{figure}[t!]
	\centering
	\includegraphics[width=.5\textwidth]{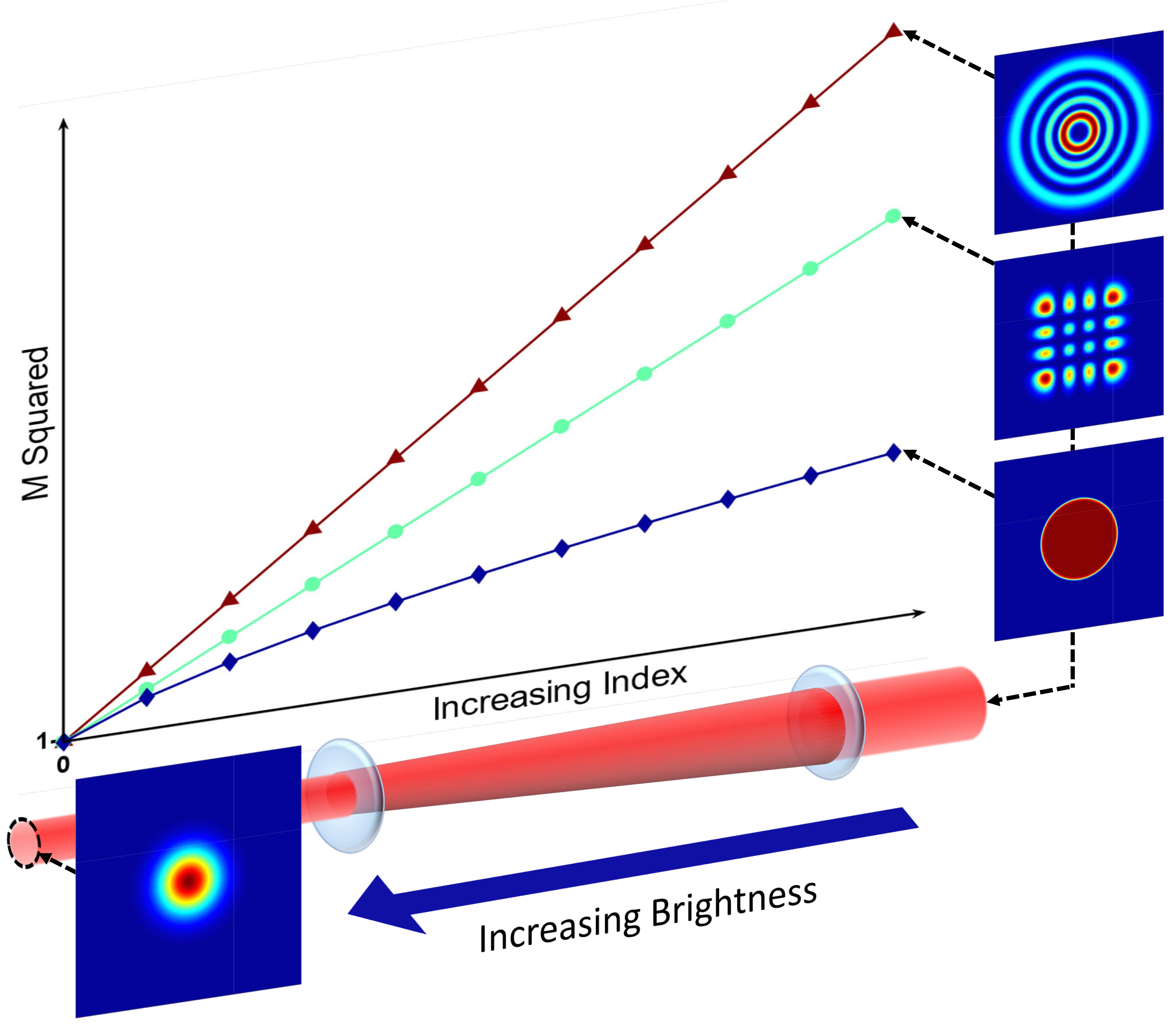}
	\caption{The beam quality of structured light fields decreases with increasing mode index, resulting in an increase in the beam quality factor, $M^2$, shown here for the examples of Hermite-Gaussian, Laguerre-Gaussian and Super-Gaussian beams.  We will show that it is possible to move ``down the curve'' towards a Gaussian ($M^2 = 1$) in a lossless manner using an optical system with two phase-only elements, thus enhancing the brightness, shown here conceptually.}
	\label{fig:indecies m squared brightness}
\end{figure}

High brightness beams are a core requirement of many optical processes \cite{ready1997industrial,shukla2012influence,shukla2015understanding}, and encompass the ability to delivery high power in a concentrated manner to a target.  While the divergence of a beam can be altered by quadratic index optical elements (e.g., lenses), its space-bandwidth product cannot. For this reason, the brightness $B$ is defined as   

\begin{equation}
\begin{aligned}
B = \frac{P}{\lambda^2(M^2)^2},
\end{aligned}
\label{eqn: brightness}
\end{equation}
\noindent where $P$ is the power per unit surface area, $\lambda$ is the wavelength and $M^2$ is the beam quality factor \cite{siegman1993defining}.  The brightness of a laser can be improved either by increasing the power ($P$), or by decreasing the $M^2$ (increasing the beam quality). Increasing the power from a laser often results in thermal phase aberrations \cite{koechner1970thermal,siegman1986lasers,innocenzi1990thermal,chenais2006thermal}, which decreases the beam quality factor \cite{koechner2013solid}.  Higher output powers can also be achieved through higher-order modes with a larger gain volume \cite{ngcobo2013exciting,bell2017excitation}, but with a reduced beam quality factor due to the higher mode index \cite{parent1992propagation,saghafi1998beam,borghi1997m,chen2011beam,siegman1998maybe,forbes2014laser}, as illustrated it Fig.~\ref{fig:indecies m squared brightness}. There are means to mitigate these effects, including coherently combining multiple independent laser sources \cite{fan2005laser,redmond2011active,montoya2012external,ramirez2015coherent}, interferometric beam correction \cite{machavariani2002efficient}, the use of correction optics \cite{haddadi2018structured,lubeigt2009reduction,gerber2006generation,lubeigt2002active} and custom laser cavities \cite{forbes2019structured,naidoo2018brightness,cagniot2011variant,hasnaoui2010properties,de2006laser,damzen1995self}, but these are complex solutions with limited general applicability.  External brightness enhancement by beam transformation has also proved limiting: it has been shown that the beam quality factor cannot be improved by binary diffractive optical elements \cite{siegman1993binary}, while demonstrations with single continuous phase elements \cite{oron2000continuous,alda2001quality} have so far resulted in a lossy or imperfect process, e.g., using a continuous spiral phase to convert a vortex beam into a Gaussian actually results in many higher-order radial modes and reduced power in the desired mode \cite{karimi2007hypergeometric,sephton2016revealing}. So the beam quality is improved but not the brightness.

Here, we show that the beam quality factor of an arbitrary beam can be improved by a phase-only transformation.  Importantly, by performing the transformation in two steps the approach is in principle lossless, hence $P$ can be maintained while the $M^2$ can be reduced, thus increasing the brightness, $B$.  The consequence is that $P$ can be maximised at the source and the beam quality corrected externally after. If an initial $M^2$ beam is converted to a Gaussian ($M^2 = 1$) while maintaining the same power (and wavelength), then the brightness can be enhanced by a factor $\eta = M^4$. While one-step approaches have been suggested to improve the beam quality factor, they come at the expense of loss or imperfect transformation, hence cannot improve the brightness.

The lossless two step approach is depicted graphically in Fig.~\ref{fig:indecies m squared brightness} for some example structured light beam types. Structured light is a topical field \cite{roadmap} and often the structure is desired.  But it can be undesirable in the context of laser brightness, e.g., if a higher-order mode is used for power extraction from a laser.  In general, the higher the mode order the higher the $M^2$ of the beam, as shown in Fig.~\ref{fig:indecies m squared brightness} for some common beam types. Here we address how to move ``down the curve'' towards lower $M^2$ without the loss of power, ideally to a Gaussian beam with the highest beam quality (lowest beam quality factor), $M^2 = 1$.  Our method is based on a conformal mapping approach \cite{bryngdahl1974geometrical,dickey2018laser} with two non-binary phase elements, illustrated in Fig.~\ref{fig:indecies m squared brightness}, and thus is not in contradiction to prior work \cite{siegman1993binary}. 

It is instructive to consider the problem in reverse.  The target beam, a Gaussian of amplitude $A(x,y)$, is passed through a phase-only optical element with a transmission function $t = \exp(i \phi)$ so that in the far field the beam is given by $u(x,y) = Q(x,y)\text{exp}(i\psi)$, where $Q(x,y)$ is some spatial profile and $\psi$ is the phase of the beam.  Say this is in fact our initial beam with structure we wish to remove.  From the reciprocity of light, if we pass it through an optical element with a phase of $-2\psi$, the resulting conjugate phase, $-\psi$, will cause the beam to ``unravel'' as it propagates, effectively destructuring it, arriving at the next phase element with the form $u(x,y) = A(x,y)\text{exp}(-i\phi)$. The phase-only element with $t = \exp(i \phi)$ now removes the unwanted phase from the beam, leaving only the Gaussian amplitude. Thus there are two phase elements needed, a first with a transmission function of $t = \exp(-i 2\psi)$ and a second with a transmission function of $t = \exp(i \phi)$. Fortunately, the calculation of the phase elements can be done, sometimes analytically, by adapting a variety of approaches \cite{bryngdahl1974geometrical,dickey2018laser,reddy2019robust,hoffnagle2000design,dorrer2007design,jiang2019free,dickey1996gaussian} depending on the implementation.
\begin{figure}[t!]
	\centering
	\includegraphics[width=.5\textwidth]{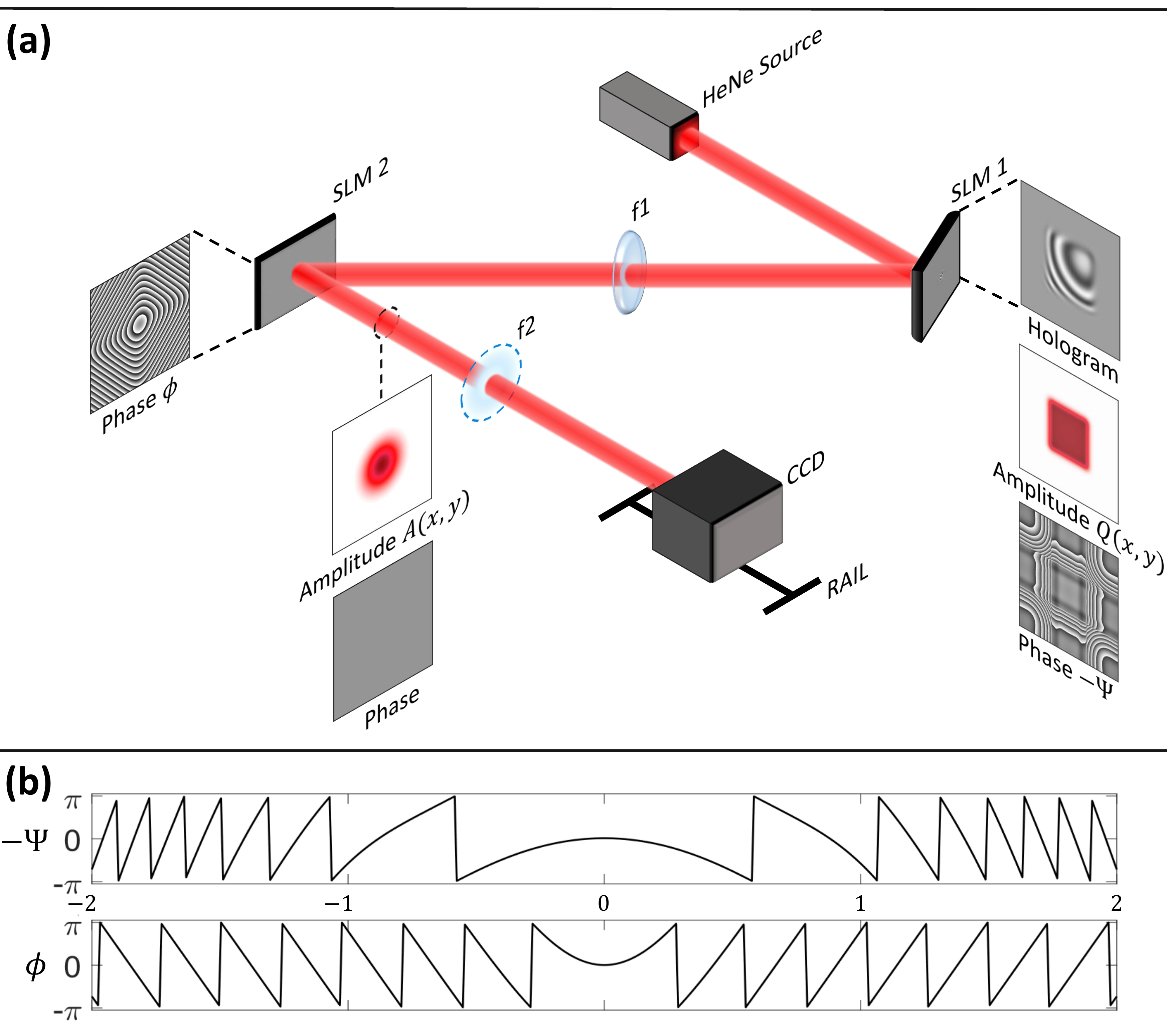}
	\caption{(a) The experimental setup consists (conceptually) of two SLMs to implement the desired phase transformations. SLM 1 acts as the first phase element creating a beam of amplitude $Q(x,y)$ with phase $-\psi$. The Fourier transform of this beam is incident on the second phase element (SLM 2) with phase $\phi$. The final beam, now having a Gaussian amplitude $A(x,y)$ and a flat phase front is imaged by a camera. (b) The phase profiles $-\psi$ and $\phi$ encoded on SLM 1 and SLM 2 respectively. The elements match the case depicted in panel (a) and are plotted on a shared transverse axis in millimeters.}  
	\label{fig:ExpSetup}
\end{figure}
\begin{table}[h!]
	\centering
	\caption{\bf The \mbox{\boldmath$M^2$} values of the initial beam profiles used in the experiment, and the predicted brightness enhancement factor, \mbox{\boldmath$\eta$}.}
	\begin{tabular}{ccc}
		\hline
		\thead{Profile\\} & \mbox{\boldmath$M^2$} & \mbox{\boldmath$\eta$} \\
		\hline
		Super-Gaussian (Radial \& Cartesian) & $3.64$ & $13.25$\\
		Radial Annulus & $16.71$ & $279.22$\\
		Cartesian Linear & $3.90$ & $15.21$\\
		\hline
	\end{tabular}
	\label{tab:m squared values}
\end{table}
\begin{figure*}[h!]
	\centering
	\includegraphics[width=\textwidth]{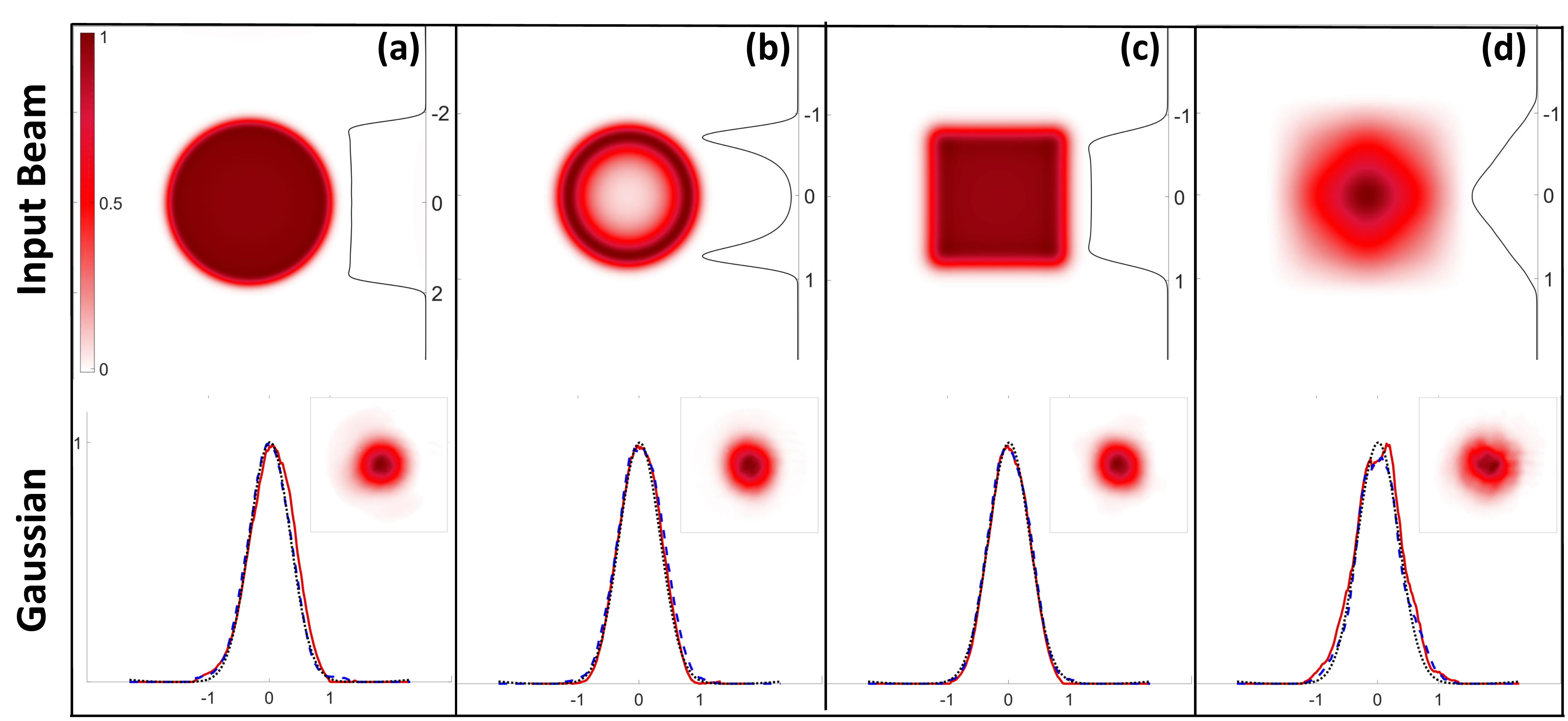}
	\caption{Input beams shown as an intensity and profile, each with a flat phase, and the measured Gaussian output beam, shown for the four tested cases: (a) Radial Super-Gaussian, (b) Radial annulus, (c) Cartesian super-Gaussian and (d) Cartesian linear, a beam with a linear intensity gradient. The examples cover Radial and Cartesian symmetry in high $M^2$ initial beams.  The Gaussian beams are shown as cross-sectional profiles for $x$ (blue dashed) and $y$ (red solid) , with the black dotted profiles the theoretical prediction.  The insets show the measured camera intensities. All profiles are plotted on an axis in millimeters.}
	\label{fig:Results}
\end{figure*}
Our implementation uses Spatial Light Modulator (SLM) to encode the phases as multilevel computer generated holograms.  We used a single device with a split-screen approach, and henceforth will explain the implementation conceptually with the two parts of the screen labelled SLM 1 and SLM 2.  It is important to note that while the SLMs are lossy devices (with an efficiency of approximately $60\%$), this implementation is purely as a proof-of-principle, which is convenient to do with re-writable holograms.  Replacing the SLMs with kinoform diffractive structures, metasurfaces, or free-form refractive optics, would make the process maintain the power, $P$ (in practice there are always small losses in such systems, but in theory the optics can be made lossless).  Using the setup shown in Fig.~\ref{fig:ExpSetup}, we first create some initial structured beam, modify it through a two-step process, and then measure the $M^2$ value of the resulting Gaussian beam. To begin, a well expanded and collimated beam from a HeNe laser was incident on SLM 1 which encoded some desired initial beam by complex amplitude modulation \cite{rosales2017shape,Forbes2016} as well as with the appropriate phase of the first transforming element.  In a real-world application the initial beam would already exist from some source so that only the transforming phase would be encoded in the first element.  The beam was then Fourier transformed to the far field and imaged to SLM 2, which applied the second phase to flattened the wavefront of the incident beam to result in a Gaussian beam with a flat phase. A Point Grey Firefly camera was used to measure the beam profiles at various positions. In particular, to measure the $M^2$ value of the final beam, lens f2 was inserted into the setup and the camera moved on a rail through the waist of the beam and the second moment width calculated from the captured images, from which the $M^2$ could be inferred. 

\begin{figure}[t!]
\centering
\includegraphics[width=.5\textwidth]{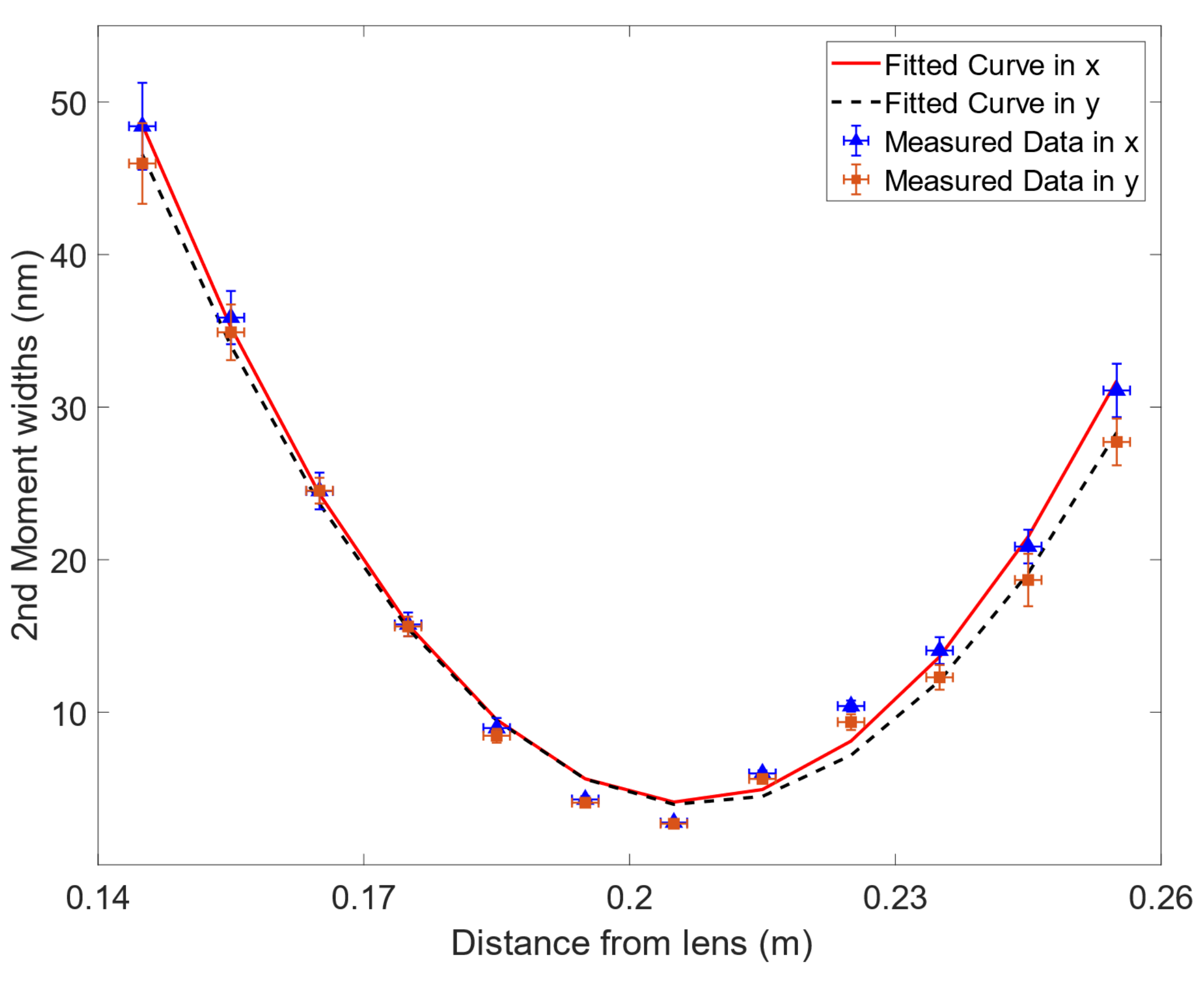}
\caption{The $2^\text{nd}$ moment width as a function of propagation distance from the focusing lens, shown as measured data and fitted curves.  This example plot is for the Radial Super-Gaussian case.}
\label{fig:m squared curve}
\end{figure}

To test the concept, we used as initial amplitude profiles those given in Table~\ref{tab:m squared values}, each with a flat initial phase.  The table shows their $M^2$ values as well as the theoretical brightness enhancement that would be expected in a lossless implementation.  The measured beams are shown in Fig.~\ref{fig:Results}, where we find excellent conversion from the initially structured beams to a Gaussian beam for all cases, with correlation factors in excess of 98\%.  The Gaussian nature is further corroborated by a full propagation analysis, shown in Fig.~\ref{fig:m squared curve} for the radial Super-Gaussian example. We find a beam quality factor decrease from $M^2 = 3.64$ to $M^2 = 1.06 \pm 0.05$, yielding a brightness enhancement of $\eta = 11.79 \pm 1.20$ when ignoring the SLM losses, in close agreement with the prediction of $\approx 13$.
The near perfect correlation between the measured and predicted Gaussian beams shows the viability of the method - we can improve the beam quality factor significantly, and hence enhance the brightness significantly.  This is not in contradiction to the prior argument that the beam quality could not be improved by binary diffractive optics \cite{siegman1993binary}.  In that scenario, the binary optics were envisaged to flip the phase jumps in higher-order modes and by propagation convert them to Gaussian-like beams in the far field. Indeed this does not improve beam quality or brightness.  Instead, we employ a two-step approach to conformally transform one beam into another, selecting the final beam to be Gaussian.  Thus both the principle and execution differ from Ref.~\cite{siegman1993binary}.  Importantly, our approach is sufficiently general so as to be applicable to beams with novel vectorial characteristics \cite{zhan2009cylindrical,rosales2018review} by implementing the solution with meta-surfaces or by using the present approach on each polarization component separately.  Finally, we remind the reader that for convenience we have used SLMs, exploiting their re-writable functionality.  For a true lossless implementation there are many options available, e.g., free-form optics as refractive or diffractive solutions. 

To conclude, we have revisited an old problem and demonstrated a simple and scaleable approach to enhancing brightness of laser beams, particularly the problematic case of converting higher-order modes to the lowest order mode (Gaussian).  We have demonstrated the concept experimentally and shown that a significant brightness enhancement is possible, up to $M^4 \times$, i.e., the higher the order of the initial beam the better the enhancement.  This work will be relevant to many high-power laser applications, for instance, improving the brightness of diode laser bars or high-power solid state lasers. 

\section*{Funding}We thank the Council of Scientific and Industrial Research with the Department of Science for the funding provided through the Interbursary Incentive Funding Programme (CSIR-DST IBS).
\section*{Disclosures}The authors declare no conflict of interest in the production or publication of this work.
\bigskip




\end{document}